\documentclass[aps,twocolumn,amsmath,amssymb]{revtex4}

\usepackage{graphicx}
\usepackage{url}

\usepackage[ pdftex, plainpages = false, pdfpagelabels,
         pdfpagelayout = useoutlines,
         bookmarks,
         bookmarksopen = true,
         bookmarksnumbered = true,
         breaklinks = true,
         linktocpage=all,
         pagebackref=false,
         colorlinks = true,
         linkcolor = BrickRed,
         urlcolor = blue,
         citecolor = BrickRed,
         anchorcolor = green,
         hyperindex = true,
         hyperfigures
         ]{hyperref}

\usepackage{cancel}
\usepackage[dvipsnames,tables]{xcolor}

\makeatletter
\g@addto@macro{\UrlBreaks}{\UrlOrds}
\makeatother

\begin{document}
\title{A Possible Link Between Pyriproxyfen and Microcephaly}
\author{Dan Evans, Fred Nijhout$^\dag$, Raphael Parens$^*$, Alfredo J. Morales$^*$ and Yaneer Bar-Yam$^*$}
\affiliation{ $^*$New England Complex Systems Institute, 210 Broadway Suite 101, Cambridge MA 02139,  $^\dag$Department of Biology, Duke University, Durham, NC 27708} 
\date{April 13, 2016}
\begin{abstract}
The Zika virus is the primary suspect in the large increase in incidence of microcephaly in 2015-6 in Brazil, however its role is not confirmed despite individual cases in which viral infections were found in neural tissue. Here we consider the alternative possibility that the use of the insecticide pyriproxyfen for control of mosquito populations in Brazilian drinking water may actually be the cause.
Pyriproxifen is a juvenile hormone analog which has been shown to correspond in mammals to a number of fat soluble regulatory molecules including retinoic acid, a metabolite of vitamin A, with which it has cross-reactivity and whose application during development has been shown to cause microcephaly. 
Methoprene, another juvenile hormone analog that was approved as an insecticide based upon tests performed in the 1970s, has metabolites that bind to the mammalian retinoid X receptor, and has been shown to cause developmental disorders in mammals. 
Isotretinoin is another example of a retinoid causing microcephaly in human babies via maternal exposure and activation of the retinoid X receptor in developing fetuses.
Moreover, tests of pyriproxyfen by the manufacturer, Sumitomo, widely quoted as giving no evidence for developmental toxicity, actually found some evidence for such an effect, including low brain mass and arhinencephaly---incomplete formation of the anterior cerebral hemispheres---in exposed rat pups. Finally, the pyriproxyfen use in Brazil is unprecedented---it has never before been applied to a water supply on such a scale. Claims that it is not being used in Recife, the epicenter of microcephaly cases, do not adequately distinguish the metropolitan area of Recife, where it is widely used, and the municipality, where it is not. Given this combination of information about molecular mechanisms and toxicological evidence, we strongly recommend that the use of pyriproxyfen in Brazil be suspended until the potential causal link to microcephaly is investigated further. 
\end{abstract}
\maketitle

\section{Overview}
In 2015 and early 2016 over 6,000 suspected cases of microcephaly and other neurodevelomental disorders have been reported in Brazil, primarily in the northeast, a dramatic increase over expected numbers \cite{MdS2016_2}. 
The cause of these developmental disorders has been widely attributed to a Zika virus epidemic first detected in May, 2015 \cite{WHO2015}. While evidence exists, particularly in neurological infections of a few microcephalic cases \cite{Mlakar2016,Tang2016}, only $12\%$ of confirmed microcephaly cases have also been confirmed as having Zika infections \cite{MdS2016_2}, preventing health authorities from determining a conclusive causal link \cite{schuler2016,Parens2016}. The strongest potential counter evidence is present in the geographic distribution of cases, including an absence of cases in Colombia. In the first 12 full weeks of 2016, 34 cases of microcephaly have been identified in Colombia \cite{Poho8April2016,WHOApril72016}, compared to 32 cases expected based upon 140 annual background cases, and not indicative of the two orders of magnitude larger number of cases found in northeast Brazil. 
Of the 34 cases of microcephaly 7 have been reported to have positive Zika tests \cite{WHOApril72016} with the rest under investigation.  
Another widely reported case in Panama \cite{Panama2016} is consistent with background rates. Reporting of the Colombia Zika outbreak began 
in October \cite{WHO2015}.  If Zika is the cause many more cases will be appearing shortly. If more cases of microcephaly are not reported by June/July of 2016 then this would be evidence against the causal link of Zika to microcephaly \cite{Cobb2016}. 
A prior outbreak in French Polynesia has been retrospectively analyzed \cite{Cauchemez2016}. The best fit model implies increased incidence of microcephaly resulting from maternal exposure in the first trimester of pregnancy, providing strong statistical evidence, though the few cases (seven during the critical period corresponding to a $1\%$ incidence) may not be conclusive, and the interpretation is complicated by the inclusion of still-births and aborted fetuses. Including just live births would yield only one microcephaly case over background. 
A preliminary cohort report on outcomes of Zika infected pregnant women in Rio de Janeiro appears to provide strong evidence \cite{Brasil2016}, however, two of the four cases in which microcephaly is reported through ultrasonographic features are associated to infections in the 5th and 6th months of pregnancy, inconsistent with both the French Polynesian analysis showing primarily early pregnancy effects and with the absence of reported cases in Colombia. Indeed, while 4 out of 23 pregnancies are reported as having anomalous ultrasounds when exposures happened in the first 4 months, 3 out of 9, (5 out of 9) pregnancies reported as infected in the 5th (6th or later) month have anomalies, including 2 still births in addition to the microcephalic ultrasounds \cite{error}. If similar exposure impacts were present in Colombia, a large number of anomalies, still births and microcephalies would already be reported. The seemingly inconsistent two late term microcephaly cases were eventually reported as small for gestational age rather than microcephaly at birth. While this might help resolve some of the discrepancy with Colombia, discounting the ultrasonographic findings leaves only one live microcephaly birth in the report. Finally, the Zika infected and small uninfected populations of the study are low/high income biased and therefore may also be to environmental exposures of teratogenic agents associated with mosquito control that correlate with areas that have higher likelihood of infection. Perhaps most puzzling, the reported tracking of microcephaly through ultrasound during pregnancy in these two studies seems inconsistent with the absence of reports in Colombia. Thus, while it is possible that a complete picture consistent with Zika as the cause can be constructed with available information, questions remain and full confirmation is not yet possible. Alternative hypotheses about causes of microcephaly should still be considered. 

A group of physicians in Argentina and Brazil have suggested that widespread use of the pesticide pyriproxifen \cite{Pubchem} to reduce mosquito populations because of a dengue epidemic in 2014 may be the cause of microcephaly \cite{Physicians2016}. This suggestion has been challenged both by authorities based upon a claim of lack of evidence \cite{Costa2016,Debunk} and by skeptics of conspiracy theories \cite{Gorski2016}. We have previously reviewed the primary evidence \cite{Parens2016}, including evidence in rat toxicology studies performed by its manufacturer Sumitomo \cite{Saegusa1988}, concluding that pyriproxifen is a credible candidate so that further inquiry is warranted. 

Here we discuss the molecular mechanisms associated with pyriproxyfen in insect and mammalian development and summarize toxicological studies and insecticide use in Brazil. Of particular note is that pyriproxyfen, a biochemical analog of juvenile hormone in insects, is cross-reactive with the retinoic acid / vitamin A regulatory system of mammals, and that exposure to retinoic acid has been shown to cause microcephaly \cite{Encyclopedia}. Existing toxicological studies and experience with public use do not provide evidence to counter this causal chain. Toxicological studies, if anything, provide evidence for neurodevelopmental toxicity at a level consistent with that found for microcephaly in Brazil. The use of pyriproxyfen in the water supplies in Brazil is at an unprecedented scale and neither prior studies nor experience is able to provide information about its effects. Our review of the currently available information points to both adequate reasons to suspect this insecticide as a developmental disruptor causing microcephaly and inadequate testing of its toxicology or use in water supply. 

\section{Potential causes and timeline}

The World Health Organization (WHO) identifies the most common causes of microcephaly as \cite{WHO2016}: infections, exposure to toxic chemicals, genetic abnormalities, and severe malnutrition while in the womb. The first includes toxoplasmosis, which is caused by parasites in undercooked meat, rubella, herpes, syphilis, cytomegalovirus and HIV. The second includes exposure to heavy metals, such as arsenic and mercury, as well as alcohol, smoking and radiation exposure. The third may include problems such as Down syndrome. While the first category points to the possibility of infectious causes, the second points to the possibility of chemical ones.

The suspected association of increased microcephaly cases with Zika and pyriproxyfen relies first on an increase in the cause with the increase in number of cases. The outbreak of the former was recognized in May 2015, and the use of the latter in the fourth quarter of 2014, with the incidence of microcephaly cases beginning in October of 2015. The precise dates of the increase in cases are believed to be uncertain because of potential problems with underreporting prior to the medical alert and overreporting afterwards \cite{schuler2016}. 

\section{Pyriproxyfen molecular mechanisms}

Juvenile hormone and retinoic acid are both lipid-soluble terpenoids.  They act as signaling molecules that control a broad diversity of embryonic and postembryonic developmental processes in insects and vertebrates, respectively.  Juvenile hormone is best known for its role in the control of metamorphosis and reproduction in insects \cite{Riddiford2012, Wheeler2003}, whereas retinoic acid is involved in the development of the nervous system in vertebrates \cite{Rhinn2012}.  

These two classes of hormone-like molecules share some molecular similarities and are capable of some degree of cross-reactivity.  For instance, retinoic acid is known to mimic some of the effects of juvenile hormone when injected into insects \cite{Nemec1993}, and juvenile hormone and its analogs are known to bind to the vertebrate retinoic acid receptor \cite{Palli1991, Jones1995, Harmon1995}.  It is possible therefore that pyriproxifen, a powerful juvenile hormone analog \cite{Dhadialla1998}, can bind to the retinoic acid receptor.  When it does so it could either activate the receptor at inappropriate times in development, or act as a blocker that prevents the normal retinoic acid from binding to the receptor when needed.  The retinoic acid receptor normally turns on gene expression in development, so either an inappropriate activation, or an inhibition at a critical time, could be expected to lead to developmental abnormalities. 

More specifically, another juvenile hormone analog that has been approved as an insecticide, methoprene, has also been shown to have metabolites that bind to the mammalian retinoid X receptor, and has been shown to cause developmental disorders in mammals \cite{Harmon1995,Unsworth1974}. 

Isotretinoin is a retinoid that is widely used in medicine, but counter indicated in women who are pregnant or might become pregnant. It causes microcephaly in human babies via maternal exposure and activation of the retinoid X receptor in developing fetuses \cite{Stem1989,Irving1986}.

The impact of retinoids on abnormal development (teratogenesis) has been demonstrated to be sensitive to genotype, developmental stage of exposure, and to result in death, malformation, growth retardation,
and/or functional disorder \cite{Collins1999}. The juvenile hormone itself and different juvenile hormone analogs have different binding properties to mammalian retinoic acid receptors, which could therefore produce different effects, including teratogenicity. These effects remain poorly understood \cite{Wheeler2003,Flat2006}.

\section{Toxicological Studies}

The potential link between pyriproxyfen and microcephaly has been challenged based upon the existence of toxicological studies by the manufacturer Sumitomo. However, we have reviewed their studies and find them largely restricted to analysis of the impact on adult animals. In the experiments on developmental toxicity, the ability to identify a link to microcephaly is weak based upon the specific tests that have been performed. Rather than measurements, in most tests visual macroscopic observation of rat and rabbit fetuses or pups/kits were used \cite{sumitomo}, for which standards of microcephaly determination may not be sufficiently well established or comparable across species. 
Still, in the most relevant experiment there is a reported test of brain weight and neurodevelopmental disorders in rat pups, and in that study, there is actual evidence for microcephaly.

Specifically, the most relevant study to a determination of neurodevelopmental toxicity \cite{Saegusa1988} considered brain and behavioral effects of rat pups exposed to pyriproxyfen during days 7 to 17 of gestation, which lasts 21 days. 
The experimental group of 36 pregnant rats in each of four test groups were fed dosage levels of 0, 100, 300 and 1000 mg/kg/day. The pups were checked for physiological deformations and organs weighed. 

From each dosage level litters of pups were obtained. For 99 pups in the 100 mg/kg and 78 pups of the 1000 mg/kg dosage groups no relevant developmental disorders were reported. Of the 99 pups in the 300 mg/kg dosage 1 (1\%) had Arhinencephaly and 1 (1\%) had Thyroid hypoplasia, The former would be consistent with concerns about neurodevelopmental disorders of the type of microcephaly.

Of the resulting offspring, 2 male and 2 female per dosage were kept alive for emotional/mental testing at 4 and 6 weeks of age and their brains were subsequently weighed at 8 weeks. One of the groups, the males of the 300 mg/kg group, had statistically significant lower brain weight at 8 weeks, implying that at least one of the only 12 pups tested in this way had substantially reduced brain weight.

These tests provide evidence that microcephaly may be an outcome of the application of pyriproxyfen to rats and other mammals. The conventional toxicology assumptions of dose dependence used in interpreting these results  infer from the absence of similar observations in the 1000 mg/kg dosage group that the cases in the 300 mg/kg group are not relevant \cite{sumitomo}. This assumption may not be reliable due to genetic variability, regulatory sensitivity, and limited numbers of observations. Instead, the outcomes might be interpreted as an estimated probability of incidence of 1 in 12. Considering the differences between humans and rats, the values for humans may be higher or lower than this. Note that the incidence of microcephaly in Brazil is estimated to be 1\% \cite{MdS2016_2}. The need for careful studies is apparent in that tetragonicity of a retinoid varies widely across mammalian species, and this variation varies across different retinoids \cite{Wilhite1986, Irving1986, Tzimas1994, Nau1993, Eckhoff1997}. Regulatory systems are remarkably sensitive to small molecular densities. Given the low incidence of human cases of microcephaly in affected populations, experiments that are designed to test for microcephaly in mammals, at the very least, should be designed to identify incidence at this level. Moreover, they should be unambiguously capable of detecting microcephaly. 

Finally, there are multiple drugs that passed conventional regulatory animal testing and are now known to be linked to microcephaly, including phenytoin and metotrexate \cite{Homes2001, Hyoun2012, swetox}, so that assurances based solely upon conventional regulatory testing are inadequate.

\section{Use of Pyriproxyfen in water supply in Brazil}

Pyriproxyfen has been used as an insecticide in northeast Brazil in response to an outbreak of dengue starting in the fourth quarter of 2014 \cite{Reis2016}. It is applied primarily to water storage containers used for home drinking water in areas that do not have a municipal water supply. 

The potential link between pyriproxyfen and microcephaly has been challenged based upon its use in other parts of the world. Pyriproxyfen is an insecticide that has received widespread regulatory approval \cite{EPA,WHO2004}. It is used on agricultural crops \cite{Devillers2013}, and against insects in households and pets, e.g., on pet collars \cite{NRDC2000}. The use of pyriproxyfen in the water supply in Brazil, however, does not have precedent. It was tested in a few communities in Malaysia \cite{Invest2008}, Peru \cite{Sihuincha2005}, Colombia \cite{Overgaard2012} and Cambodia \cite{Seng2008}. These were small case studies and only Cambodia and Colombia included use in the drinking water supply \cite{Invest2008}. Moreover, the research on impacts included measures of mosquito control but did not consider developmental effects in humans. 

Prior to 2014, Brazil primarily used temefos (an organophosphate) in water supplies to reduce mosquito populations. The switch to pyriproxyfen was made due to increasing temefos resistance in mosquitoes \cite{PdS2014}. Some areas of Brazil do not use pyriproxyfen, using BT Toxin instead. The Brazilian Ministry of Health has suggested that areas not using pyriproxyfen also report cases of microcephaly. For example, there is reported to be no use of pyriproxyfen in Recife, and two other cities of the state of Pernambuco \cite{Romo2016, Bichell2016, NoAuthor2016}, though there are cases of microcephaly and Zika. This has been cited as a compelling reason for dismissing the possibility that pyriproxyfen is a cause of microcephaly \cite{NoAuthor2016}. However, according to the Dengue control office of Pernambuco, pyriproxyfen is widely used in the Recife metropolitan area. It is not used in three specific municipalities (urban administrative areas), specifically Recife, Paulista and Jaboatao do Guararapes. Pyriproxyfen exposure is relevant for the urban favelas and nearby rural areas where it would be used in water storage containers but is not relevant to urban areas with municipal water supply systems. Urban hospitals attract populations from a large area \cite{Douglas2016} and official reporting of cases is not broken down by municipality but rather by state, so that a direct analysis of geographical distribution of exposure to pyriproxyfen is difficult in these particular cases. Thus, further analysis is needed to determine whether the large number of cases in Pernambuco state arises from areas where pyriproxyfen is used. 

\section{Conclusion}
This paper analyzes the potential causal connection between the pesticide pyriproxyfen and microcephaly, as an alternative to Zika. Pyriproxyfen is a juvenile hormone analog, which has been shown to be cross reactive with retinoic acid, part of the mammalian regulatory system for neurological development, whose application during development causes microcephaly. This causal chain provides ample justification for pursuing a careful research effort on the role of pyriproxyfen in neurodevelopmental disorders. Counter to stated claims, existing studies of neurodevelopmental toxicity by Sumitomo, its manufacturer, provide some supportive evidence for neurodevelopmental toxicity including low brain weight in rat pups. The large-scale use of pyriproxyfen in Brazil and its coincidental timing with an increase in microcephaly cases should provide additional motivation. We believe that this evidence is strong enough to warrant an immediate cessation of pyriproxyfen application to Brazilian water supplies until additional research can be carried out on its neurodevelopmental toxicity. 
Where insecticides are considered essential, Bt toxin is considered a safe alternative that has been used in water systems in Recife since 2002 \cite{BtToxin}. Other more effective vector control methods may also be used \cite{Regis2013,Regis2008,Bar-Yam2016}.

\textbf{Acknowledgements:}
We thank Audi Byrne for helpful discussions, and Keisuke Ozaki for helpful communications about Sumitomo toxicology tests.


\begin{thebibliography}{20}
\bibitem{error} There is a discrepancy between the number of reported (42) and the number of cases shown in Fig. 1 of that paper (41), the numbers we cite are those present in the figure as it is unclear which week the additional exposure is in.
\end{thebibliography}
\end{document}